\documentclass[12pt,preprint]{aastex}

\newcommand{\kms}{\mbox{km s$^{-1}$}}

\newcommand{\Msun}{\mbox{$M_{\sun}$}}
\newcommand{\Lsun}{\mbox{$L_{\sun}$}}
\newcommand{\skipthis}[1]{}

\shortauthors{Garay et al.}
\shorttitle{High-mass protostellar objects at 7 mm}
\slugcomment{Accepted by {\em The Astronomical Journal}}

\begin{document}



\title{VLA observations of candidate high-mass protostellar objects at 7 mm} 

\author{Guido Garay}
\affil{Departamento de Astronom\'{\i}a, Universidad de Chile,
Casilla 36-D, Santiago, Chile}

\author{Luis F. Rodr\'\i guez}
\affil{Centro de Radioastronom\'\i a y Astrof\'\i sica, UNAM,
Apdo. Postal 3-72, Morelia, Michoac\'an, 58089 M\'exico}

\and

\author{Itziar de Gregorio-Monsalvo}
\affil{European Southern Observatory, Alonso de C\'ordoba 3107,
Vitacura, Casilla 19001, Santiago 19, Chile}

\begin{abstract}

We present radio continuum observations at 7 mm made using the Very Large 
Array towards three massive star forming regions thought to be in very early 
stages of evolution selected from the sample of Sridharan et al. (2002). Emission 
was detected towards all three sources (IRAS 18470-0044, IRAS 19217+1651 and 
IRAS 23151+5912). We find that in all cases the 7 mm emission 
corresponds to thermal emission from ionized gas. The regions of ionized gas 
associated with IRAS 19217+1651 and IRAS 23151+5912 are hypercompact with 
diameters of 0.009 and 0.0006 pc,
and emission measures of $7.0\times10^8$ and $2.3\times10^9$ pc cm$^{-6}$, 
respectively. 

\end{abstract}



\section{Introduction}

One of the key challenges at present in the field of star formation is to
understand how massive stars (M $>$ 10 \Msun) form.
Our present understanding of star formation is primarily based on the 
observations of low-mass stars. In the current paradigm (Shu, Adams, \& Lizano 
1987; Shu et al. 1993) the formation of low-mass stars is characterized by 
an accretion phase, in which a central protostar and a circumstellar disk form 
surrounded by an infalling envelope of dust and gas, followed by a phase 
in which the protostar deposits linear and angular momentum, and mechanical 
energy into its surroundings through jets and molecular outflows.
Although this paradigm has been very successful in explaining what is observationally
known about the formation of low-mass stars (e.g, Lada 1991), its applicability to
the formation of massive stars remains arguable.
In particular, alternative mechanisms such as the merging of lower mass
protostars to form a massive protostar have received serious consideration
lately (e. g. Bonnell, Bate, \& Zinnecker 1998; Portegies Zwart et al. 1999), 
and there is evidence of strong dynamical interaction and possibly merging 
deduced from the proper motions of some young massive stars in the
Orion KL/BN region (G\'omez et al. 2005).

Whether the massive protostars are formed by accretion or merging
remains controversial. If massive O stars are formed by accretion, we expect
that disks and jets will be present in their earliest stages of evolution,
as in the case of low mass stars. Due to their large luminosities, hypercompact 
HII regions are also expected to be formed while the protostar is still undergoing 
accretion (Keto 2002, 2003). On the other hand, if formed by random coalescence 
of lower-mass stars neither disks, jets or hypercompact HII regions are expected. 
Thus, the search for jets, disks and hypercompact HII regions toward massive YSOs 
is crucial to understand their formation process.

The presence of jets and disks toward massive YSOs still lacks solid grounds.
Collimated outflows and/or disks have been found in a handful of B-type protostars:
IRAS 18162-2048 (${\cal L}\sim1.7\times10^4$ \Lsun; Mart\'\i , Rodr\'\i guez, \& 
Reipurth 1993); Cepheus A HW2 (${\cal L} \sim1\times10^4$ \Lsun; Rodr\'\i guez et 
al. 1994); IRAS 20126+4104 (${\cal L} \sim1.3\times10^4$; Cesaroni et al. 1997); 
G192.16-3.82 (${\cal L}\sim3\times10^3$; Shepherd et al. 1998; Devine et al. 1999; 
Shepherd, Claussen, \& Kurtz 2001); AFGL~5142 (${\cal L}\sim3.0\times10^3$
\Lsun; Zhang et al. 2002). None of these objects exceeds a bolometric luminosity of
$2\times10^4$\Lsun~and are thus B0 ZAMS or lower luminosity objects.
The most luminous O-type protostars associated with a jet and a collimated 
molecular outflow is IRAS~16547$-$4247, a luminous infrared source with a 
bolometric luminosity of $6.2\times10^4$\Lsun (Garay et al. 2002, 2007;
Rodr\'\i guez et al. 2005).  There is, however, no known unquestionable
case of a circumstellar disk associated with a massive O-type protostar. 
The search for disks around independent high-mass protostars is inherently difficult. 
The photoevaporation timescale of disks by the radiation from the central star
(Yorke \& Welz 1996; Hollenbach 1997) can be relatively short, and disks may
be destroyed before the dispersal of the molecular cloud core.
If other OB stars in the region have been formed before, the object studied 
may be embedded in bright free-free emission that will difficult the search 
for the relatively weak mm emission from the disk.

We report here 7 mm continuum observations, made with the Very Large Array,
towards three luminous ($L \geq 6 \times 10^4~L_\odot$) objects,  
selected from the sample of high-mass protostellar candidates of Sridharan at 
al. (2002), that are associated  with powerful bipolar CO outflows (indicating the 
presence of a collimated outflow and thus suggesting the presence of a disk).
The main goal of these observations was to search for either circumstellar disks 
or hypercompact regions of ionized gas around O-type protostars.


\section{Observations}

The observations were made with the Very Large Array (VLA) of the National
Radio Astronomy Observatory
\footnote{
The National Radio Astronomy Observatory is a facility of the
National Science Foundation operated under cooperative agreement by
Associated Universities, Inc.},
on 2003 April 25. 
Each source was observed for 45~minutes, under good weather conditions,
in the Q (7 mm) band using the standard VLA continuum mode (4IF, 50 MHz per IF).
To minimize the effect of atmospheric phase noise, we observed in the 
fast-switching mode with a cycle of 120 seconds.
The data were edited and calibrated by applying the complex gain solution from
the calibration source following the standard VLA procedures. The flux density
scale was determined by observing the source 3C286 for which we assumed flux
densities of 1.45 Jy at 7 mm. 
In order to correct the amplitude and phase of the interferometer data 
for atmospheric and instrumental effects, we observed before and after every 
on-source scan the calibrator J1851+005.
Standard calibration and data reduction were performed using AIPS. Maps were 
made by Fourier transformation of the interferometer data using the AIPS 
task IMAGR. The noise level in the images was in the range 0.12-0.16 mJy
beam$^{-1}$.  The array was in the D configuration, 
resulting in a synthesized beam of $\sim2\arcsec$ at 43.4 GHz.
In this observing session the source IRAS 18566+0408 was also observed,
finding that its 7 mm emission probably arises from dust. This
source is discussed in detail in Araya et al. (2007).

We also made use of the archival VLA radio continuum data and found that all 
three regions have been previously observed at cm wavelengths. A summary of the 
archive data used here is given in Table \ref{tbl-vlaarc}. 


\section{Results and discussion}

We detected 7 mm continuum emission towards each of the three observed sources.
The position, flux densities, and angular size of
the detected radio sources are given in Table~\ref{tbl-obspar}.
In what follows we present the results of our continuum observations
and discuss the nature of the detected sources in each of the regions,
individually. For the derivation of physical parameters we used the distances
given by Sridharan et al. (2002) and listed in Table 3. 

\subsection{IRAS 18470-0044}

\begin{figure}
\epsscale{1.0}
\plotone{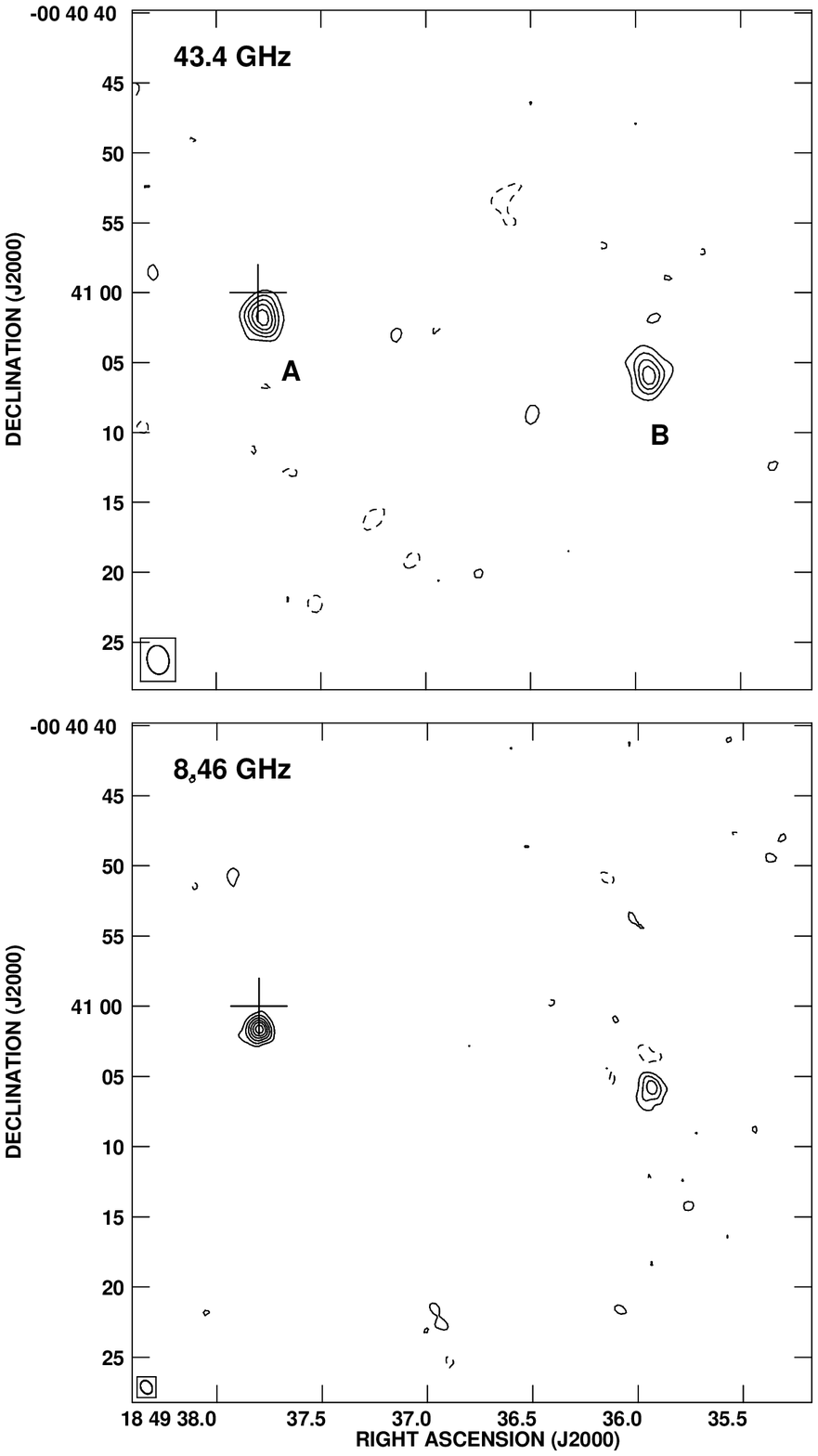}
\vspace{-20.0mm}
\caption
{VLA radio continuum maps towards IRAS 18470-0044. 
Beams are shown in the lower left corner of each panel.
Top : 43.4 GHz map. Contour levels are -1, 1, 2, 3, 4 and 5 $\times$ 0.25
mJy per beam.  Beam is $2.1\times1.6$ \arcsec. 
Bottom : 8.46 GHz map. Contour levels are -1, 1, 2, 3, 4, 5 and 6 $\times$ 
0.20
mJy per beam. Beam is $1.0\times0.77$ \arcsec.
The cross indicates the peak position of the dust core (Williams et al. 2004). 
\label{fig-18470maps}}
\end{figure}

Figure~\ref{fig-18470maps} shows a contour map of the 7 mm continuum emission 
observed towards IRAS 18470-0044. The emission arises from two distinct compact 
sources separated by 28\arcsec\ (labeled A and B). Both sources were also 
detected at 3.6 cm and 6.0 cm wavelengths. A contour map of the 3.6 cm emission 
is shown in the lower panel of Fig.~\ref{fig-18470maps}. Component A is 
associated with a massive dust core detected at 850 and 450 
$\mu$m by Williams et al. (2004) and at 1200 $\mu$m by Beuther et al. (2002a). 
At 1.2 mm the core has major and minor axis of 17 and 12\arcsec, respectively,
implying a core radius, estimated from the geometric mean of the axis, 
of 0.29 pc (assuming a distance of 8.2 kpc). The mass of the core determined from 
the 850 $\mu$m observations is 720 \Msun. The peak position of the core is marked 
with a cross in Fig.~\ref{fig-18470maps}.

The spectral index of the emission between 5.0 and 43.4 GHz are $-0.1\pm0.1$ 
and $-0.1\pm0.1$ for the east (A) and west (B) components, respectively,
indicating that the emission in this frequency range is optically thin 
thermal emission. We conclude that the emission from these two objects 
is free-free emission from ionized gas. Table~\ref{tbl-derived} lists
the distance [col.~(2)], the derived parameters of the regions of ionized gas:
diameter [col.~(3)],  emission measure [col.~(4)], and electron density [col.~(5)], 
the minimum number of ionizing photons required to maintain the ionization of the nebula 
[col.~(6)], and inferred spectral type of the exciting star [col.~(7)]. They were 
calculated following the formulation of Mezger and Henderson (1967), assuming that 
the gas has constant electron density and an electron temperature of $10^4$~K.

The regions of ionized gas have radii of 0.021 and 0.029 pc, in the range of those 
of UC HII regions, and densities of $4.3\times10^3$ and $2.5\times10^3$ cm$^{-3}$, 
much lower than those of UC HII regions. We suggest that these small regions of 
ionized gas are deeply embedded within molecular cloud cores with high densities 
and large turbulent motions, and have already reached pressure equilibrium with the 
dense ambient gas (e.g., De Pree, Rodr\'\i guez, \& Goss 1995; Xie et al. 1996).
The equilibrium radius is given by (see Garay \& Lizano 1999),
$$ R_f =  0.034
\left({{N_{\rm u}}\over{3\times10^{48}~{\rm s}^{-1}}}\right)^{1/3}
\left({{4\times10^{5}~{\rm cm}^{-3}}\over{n_o}}\right)^{2/3}
\left({{T_e}\over{10^4 K}}\right)^{2/3}
\left({{3~ \kms}\over{\Delta v}}\right)^{4/3} ~~{\rm pc}, \eqno(1)$$
where $n_o$ is the density of the ambient gas, $N_{\rm u}$
is the rate of ionizing photons emitted by the exciting star,
and $\Delta v$ is the line width of the molecular emission from the ambient gas.
The time to reach pressure equilibrium is
$$ t_{eq} = 4.8\times10^3
\left({{N_{\rm u}}\over{3\times10^{48}~s^{-1}}}\right)^{1/3}
\left({{4\times10^{5}~{\rm cm}^{-3}}\over{n_o}}\right)^{2/3}
\left({{T_e}\over{10^4 K}}\right)^{2/3}
\left({{3~ \kms}\over{\Delta v}}\right)^{7/3}~~{\rm yrs}. \eqno(2)$$

Component A is projected right at the center of the massive and dense core 
which has an average molecular gas density of $1.4\times10^5$ cm$^{-3}$ and 
a line width in optically thin molecular transitions of 2.8 \kms\
(Beuther et al. 2002a). Using these values and the total number of ionizing photons 
per second needed to excite this component, of 2.1$\times10^{46}$ s$^{-1}$, 
we find that the equilibrium radius is $\sim0.014$ pc, in good agreement with
the observed radius of 0.021 pc.  The time to reach pressure equilibrium is
$\sim2\times10^3$ yrs. The actual age of the IRAS 18470-0044 massive star 
forming region may be much larger than this value and thus might not correspond 
to an object in a very early stage of evolution.

The total far-infrared luminosity of the IRAS 18470-0044 source, 
whose angular extent encompasses both regions of ionized gas,
is $7.9\times10^4$ \Lsun\ (Sridharan et al. 2002). On the other hand,
the total luminosity of the ionizing sources is $2.3\times10^4$ \Lsun. 
The difference in the inferred luminosities from the radio and FIR 
observations is most likely due to the presence of a group
of non-ionizing stars which account for a fraction of the 
total bolometric luminosity but produces a negligible amount of 
ionizing photons. This explanation is strongly supported 
by our analysis of unpublished archive Spitzer IRAC observations
which show the presence of several 
embedded sources within the angular extent of the IRAS source. 
However, it should be noted that the presence of
dust absorbing ionizing stellar UV photons is most probably
also contributing to the difference in the luminosities
inferred from the radio and the FIR observations. 
   
\subsection{IRAS 19217+1651}

IRAS 19217+1651 is associated with a massive dust core detected at 850 and 
450 $\mu$m by Williams et al. (2004) and at 1200 $\mu$m by Beuther et al. (2002a). 
At 1.2mm the core has major and minor axis of 18 and 15\arcsec, respectively, 
implying a core radius, estimated from the geometric mean of the axis,
of 0.42 pc (assuming a distance of 10.5 kpc). The mass of the core determined from 
the 850 $\mu$m observations is $2\times10^3$ \Msun. Beuther et al. (2002b) detected 
class II 
methanol and water masers near the peak position of the dust core, suggesting that 
this is a massive star forming region in an early stage of evolution.

\begin{figure}
\epsscale{1.0}
\plotone{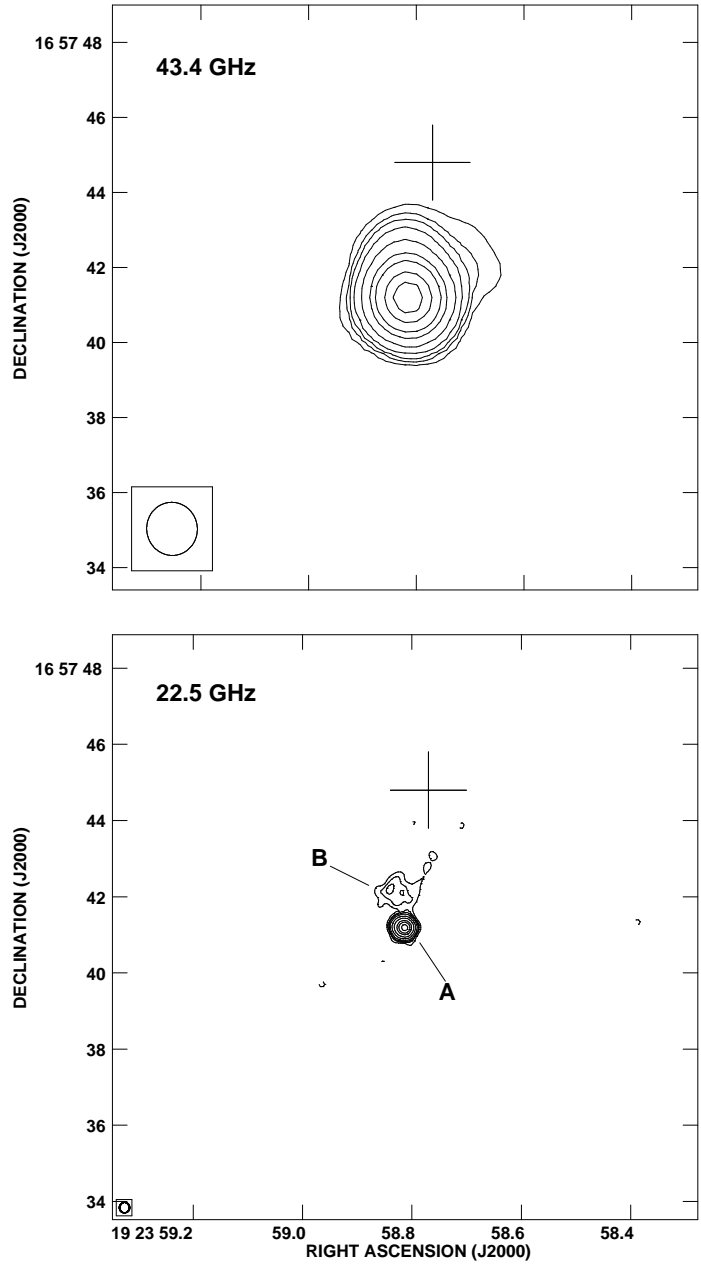}
\vspace{-30.0mm}
\caption
{VLA radio continuum maps towards IRAS 19217+1651. 
Beams are shown in the lower left corner of each panel.
Top : 43.4 GHz map. Contour levels are -1, 1, 2, 3, 5, 10, 20, 30, 50 and 70  
$\times$ 0.50
mJy per beam.  Beam is $1.42\times1.35$ \arcsec. 
Bottom : 22.5 GHz map. Contour levels are -1, 1, 2, 3, 5, 10, 20, 40, 60, 90 and 
120 $\times$ 0.30
mJy per beam. Beam is $0.27\times0.26$ \arcsec.
The cross indicates the peak position of the dust core (Williams et al. 2004). 
\label{fig-19217maps}}
\end{figure}

The upper panel of Fig.~\ref{fig-19217maps} shows a contour map of the 7 mm 
continuum emission observed with the VLA towards IRAS 19217+1651. Most of the emission 
arises from a single bright source, although there is
evidence of a weaker component to the north. VLA observations at centimeter wavelengths 
(X, U, and K bands) show that emission arises from 
two components. This is illustrated in the bottom panel of Fig.~\ref{fig-19217maps}
which presents a contour map of the emission observed at 22.5 GHz, 
showing that the emission arises from a bright compact component (labeled A), 
with an angular size of {$\sim 0.16\arcsec$}, and a more extended, weaker 
component (labeled B) located $\sim1\arcsec$ NE of component A, with an angular 
size of 0.8\arcsec. The peak of the 7 mm source coincides with component A.
The cross in both panels marks the peak position of the large dust core.

The radio continuum spectra of component A, in the range from 8.4 to 43.4 GHz, 
is shown in the upper panel of Figure~\ref{fig-spectra}. The spectra 
is reasonably well modeled (dotted line) by that of an homogeneous region of 
ionized gas with an emission measure of $7.0\times10^8$ pc cm$^{-6}$ and an 
angular size of 0.19\arcsec. The region of ionized gas is optically thick below 
15 GHz. The total number of ionizing photons per second required to excite this 
region is 5.5$\times10^{47}$ s$^{-1}$, which could be supplied by an O9.5 
ZAMS star.  The total luminosity of the region inferred from the radio 
observations is $5.4\times 10^4$ \Lsun, which is close to the luminosity 
inferred from the IRAS observations of $7.9\times 10^4$ \Lsun.
Possible explanations for the difference in luminosities is that a fraction 
of the ionizing stellar UV photons is absorbed by dust within the 
ultracompact HII region or that non-ionizing stars are present in the region
(or a combination of both). Our analysis of unpublished archive Spitzer observations
show a single bright source associated with IRAS 19217+1651 that is detected
in all four IRAC bands.  

The radio continuum spectra of component B, in the range of 8.4 to 22.5 GHz,
is consistent with a total flux density of 6.6$\pm$1.0 mJy
at 43.4 GHz, leaving a flux density of 43.8$\pm$1.0 mJy
for component A at this frequency. The flat spectrum of component B is indicative
of optically thin free-free emission. The direct interpretation of these
data is that, since the total number of ionizing photons per second required to excite this
region is 6.9$\times10^{46}$ s$^{-1}$, we are observing 
an independent HII region ionized by a B0.5 ZAMS star with a
luminosity of $2.0\times 10^4$ \Lsun. However, the high angular resolution
images (see the 22.5 GHz image in Fig. 2) suggest that component B could be
ionized gas outflowing from component A. Is this alternative possible?
The angular separation between components A and B is $\sim1{''}$
that at a distance of 10.5 kpc gives a physical distance of
$1.6\times 10^{17}$ cm. Moving at a velocity of 10 km s$^{-1}$, the ionized gas
will take about 5,000 years in moving from component A to component B.
On the other hand, since component B has a deconvolved
angular diameter of $0\rlap.{''}8$, we estimate for this
source an average electron density
of $8.9\times 10^3$ cm$^{-3}$ and a recombination timescale of only 14 years. 
Then, we conclude that component B requires its own ionizing star.

\subsection{IRAS 23151+5912}

IRAS 23151+5912 is associated with a massive dust core detected at 850 and 450 
$\mu$m by Williams et al. (2004) and at 1200 $\mu$m by Beuther et al. (2002a). 
At 1.2mm the core has major and minor axis of 16.5 and 14.4\arcsec, respectively,
implying a core radius of 0.21 pc (assuming a distance of 5.7 kpc). The mass of 
the core determined from the 850 $\mu$m observations is $3\times10^2$ \Msun. 
Water masers were detected near the peak position of the core by Beuther et al. (2002b).
   
\begin{figure}
\epsscale{1.0}
\plotone{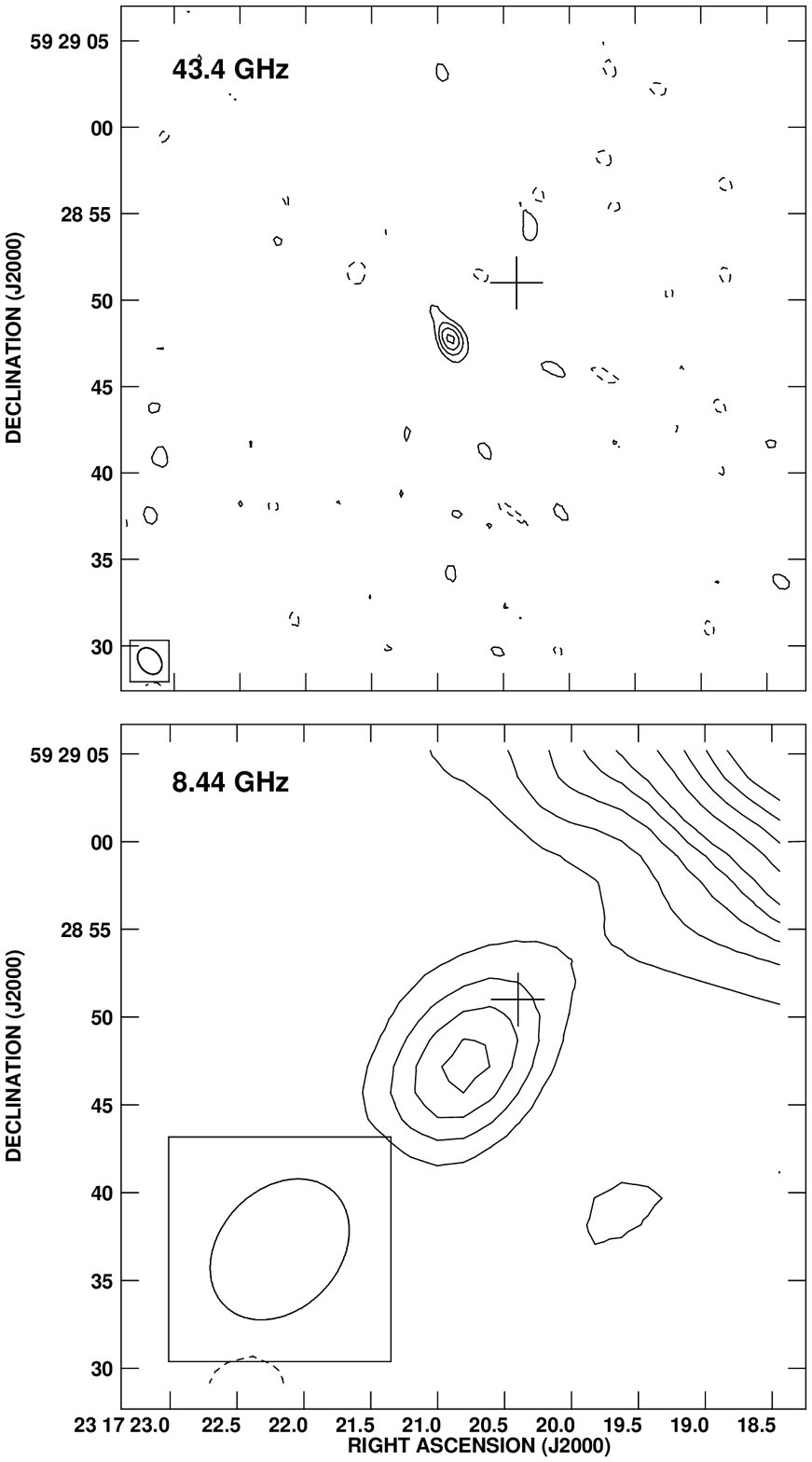}
\vspace{-30.0mm}
\caption
{VLA radio continuum maps towards IRAS 23151+5912. 
Beams are shown in the lower left corner of each panel.
Top : 43.4 GHz map. Contour levels are -1, 1, 2, 3 and 4 
$\times$ 0.40
mJy per beam.  Beam is $1.94\times1.52$ \arcsec. 
Bottom : 8.44 GHz map. Contour levels are -1, 1, 2, 3, 4, 5, 7, 9, 11 and 
13 $\times$ 0.060
mJy per beam. Beam is $9.0\times6.8$ \arcsec. 
The cross indicates the peak position of the dust core (Williams et al. 2004). 
\label{fig-23151maps}}
\end{figure}

\begin{figure}
\epsscale{1.0}
\plotone{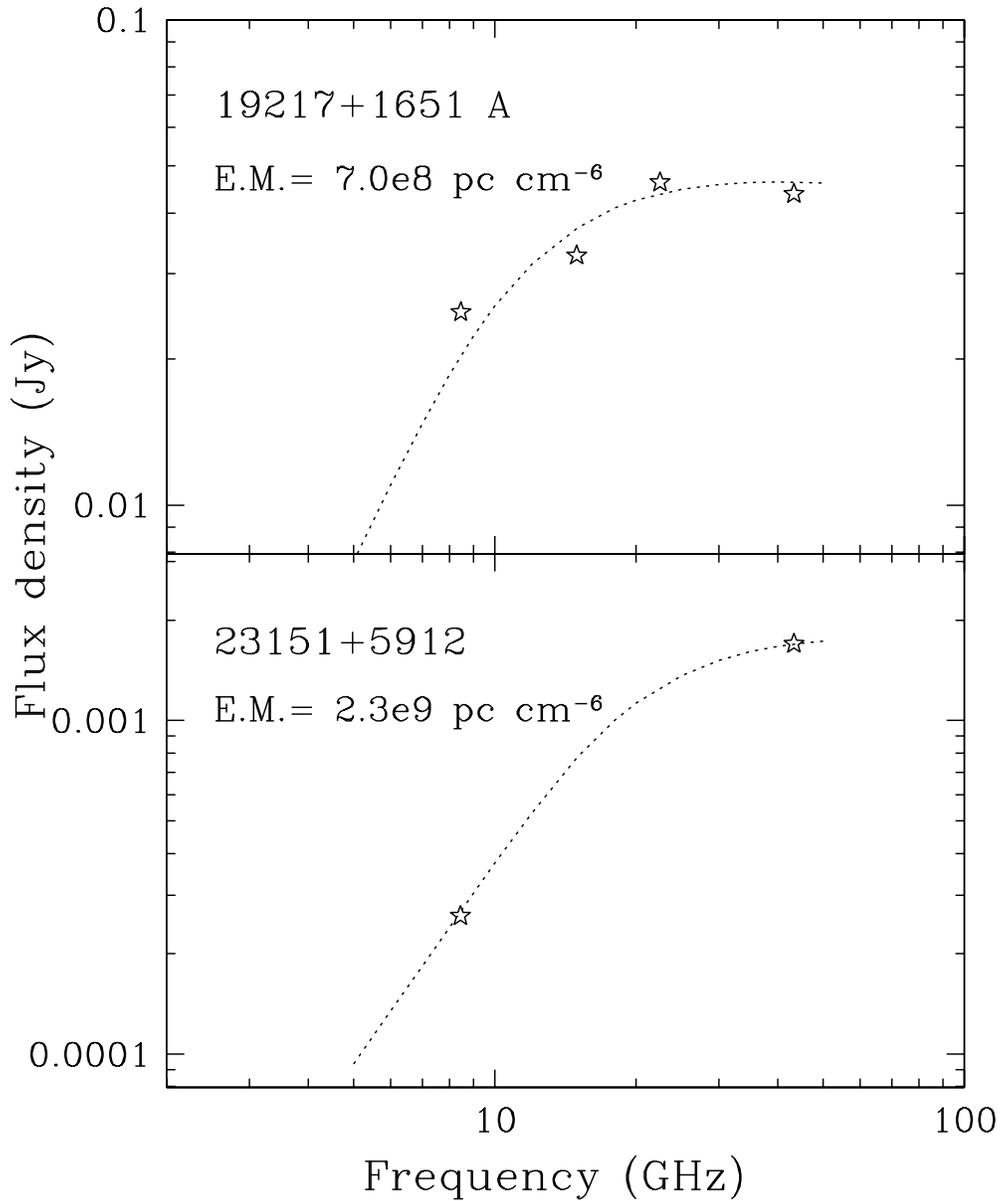}
\vspace{-30.0mm}
\caption
{Radio continuum spectra of the regions of ionized gas with high emission measures.
Dotted lines: fits to the observed spectra with a theoretical
model of an homogeneous region of ionized gas. The fitted $EM$ (in pc cm$^{-6}$)
is given in the upper right corner.
Top: Spectra of radio component A towards IRAS 19217+1651.
Bottom: Spectra of the radio source towards IRAS 23151+5912.
\label{fig-spectra}}
\end{figure}

The 7 mm continuum observations towards IRAS 23151+5912 shows the presence 
of a single source, as illustrated in the upper panel of Figure~\ref{fig-23151maps}. 
We also detected emission from this source at 3.6 cm
(with a total flux density of 0.27$\pm$0.04 mJy), as shown in the lower panel 
of Figure 3.  The radio continuum spectrum, shown in the bottom panel of 
Fig.~\ref{fig-spectra}, can be well fitted by that of an uniform density region 
of ionized gas with an emission measure of $2.3\times10^9$ pc cm$^{-6}$ and 
an angular size of 0.02\arcsec (or $6\times10^{-4}$ pc). 
The high emission measure and small size of this HII region are characteristics 
of hypercompact HII regions (Kurtz 2000, Hoare et al. 2007).
Unfortunately, the modest angular resolution of our data does not allow a direct
measurement of the angular size of this source and only an upper limit is
possible (see Table 2). Additional observations of very high angular resolution
are needed to test the nature of this source as a hypercompact HII region.
Hypercompact HII regions are thought to mark the earliest stages of 
evolution of regions of ionized gas, being formed in the accretion 
phase of hot molecular cores (Keto 2002, 2003; Gonz\'alez-Avil\'es, Lizano \& Raga 
2005; Avalos et al. 2006). We conclude that IRAS 23151+5912 is 
indeed associated with a massive star forming region in a very early stage 
of evolution.  We note that the spectral index of the 7 mm source between 8.4 
and 43.4 GHz is $1.1\pm0.2$. This value is typical of the spectral index of 
hypercompact HII regions at centimeter wavelengths, which suggests the presence 
of non-uniform density gas. With the presently available data we cannot however
discern whether or not the ionized gas exhibits density gradients.
In particular, we cannot rule out the possibility that the source
is an ionized stellar wind.

The rate of ionizing photons needed to excite the HII region is 
5.5$\times10^{45}$ s$^{-1}$, which could be supplied by an B0.5 ZAMS star
with a luminosity of $1.1\times10^4$ \Lsun. The FIR luminosity of the region 
inferred from the IRAS observations is, however, $1.0\times 10^5$ \Lsun\ 
(Sridharan et al. 2002). The most likely explanation for the discrepancy
in the luminosities inferred from the radio and FIR observations 
is that, due to the coarse angular resolution of the IRAS 
observations, the FIR luminosity includes the contribution of the star or 
stars ionizing the large cometary HII region located 30{''} NW of the dense core
(see Fig. 5). This cometary HII region has a flux density of $\sim$20 mJy,
which requires a B0.5 ZAMS star with a luminosity of
$2.0\times 10^4$ \Lsun\ to maintain its ionization. 
Another contributing effect could be absorption of UV photons by dust within 
the hypercompact HII region.  
There is no archive Spitzer data for this source.


\section{Summary}

We made 7 mm, high-angular resolution, continuum observations, using the VLA, 
towards three regions of massive star formation thought to be in early stages 
of evolution.  These regions are associated with energetic molecular outflows
and two of them were not previously detected in radio continuum 
observations with the VLA. The main objective was to search for the presence 
of disks and/or hypercompact regions of ionized gas.  Our main results and 
conclusions are as follows.

Emission at 7 mm was detected toward the three regions. Towards IRAS 18470-0044 we 
detected two objects with spectral indices between 4.8 and 43.4 GHz of $\sim-0.1$,
indicating that the emission is free-free radiation arising from optically 
thin HII regions excited by stars with spectral types B0.5. We conclude that 
these small HII regions are embedded within massive and dense cores
and have already reached pressure equilibrium with their 
dense and turbulent molecular surroundings.

In the case of IRAS 19217+1651 we detect two components: component A is
a hypercompact HII region that is optically thick
below 15 GHz, while component B is optically thin in the 
8.4 to 22.5 GHz range. We discuss if component B could be ionized gas outflowing
from component A but conclude that each component has its own ionizing star.

Towards IRAS 23151+5912 we detected a hypercompact HII region
that can be modeled as a source
with a radius of $3\times10^{-4}$ pc (or 57 AU) and an emission measure of 
$2.3\times10^9$ pc cm$^{-6}$.  We conclude that IRAS 23151+5912 is
indeed a massive star forming region in a very early stage
of evolution, with the hypercompact HII region being formed 
while the suspected hot molecular core around it is still probably undergoing accretion.

\begin{acknowledgements}

GG gratefully acknowledges support from
the Chilean Centro de Astrof\'\i sica FONDAP No. 15010003.
LFR is thankful to the support
of CONACyT, M\'exico and DGAPA, UNAM.
IdG acknowledges support from Consejer\'\i a de Innovaci\'on, Ciencia y Empresa of 
Junta de Andaluc\'\i a, grant FQM-1747.

\end{acknowledgements}


\newcommand\rmaap   {RMA\&A~}
\newcommand\aujph   {Aust.~Jour.~Phys.~}
\newcommand\rmaacs   {RMA\&A Conf. Ser.~}

\newpage

\begin{deluxetable}{lccccc}
\tablewidth{0pt}
\tablecolumns{6}
\tablecaption{VLA ARCHIVE OBSERVATIONS ANALIZED \label{tbl-vlaarc}}
\tablehead{
\colhead{IRAS source}   & \colhead{Project} &  \colhead{Observed date} 
& \colhead{Freq.} & \colhead{Conf.}  & \colhead {Synthesized} \\
\colhead{}                  & \colhead{name}      &
\colhead{}    & \colhead{(GHz)}  & \colhead{ } &
\colhead{Beam (\arcsec$\times$\arcsec)} \\
}
\startdata
18470-0044 & AH810  & 21-MAR-2005 &  4.86 &  B   & $1.96\times1.43$ \\
18470-0044 & AS643  & 02-JUL-1998 &  8.46 &  BnA & $1.01\times0.77$ \\ 
19217+1651 & AS643  & 02-JUL-1998 &  8.46 &  BnA & $1.63\times0.72$ \\ 
19217+1651 & AB1101 & 08-NOV-2003 & 14.94 &  B   & $0.40\times0.36$ \\  
19217+1651 & AB1101 & 08-NOV-2003 & 22.46 &  B   & $0.27\times0.26$ \\ 
23151+5912 & AR304  & 04-NOV-1993 &  8.44 &  D   & $8.98\times6.82$ \\ 
\enddata
\end{deluxetable}

\begin{deluxetable}{lcccc}
\tablewidth{0pt}
\tablecolumns{5}
\tablecaption{OBSERVED PARAMETERS OF 7 mm EMISSION \label{tbl-obspar}}
\tablehead{
\colhead{Radio source}     & \multicolumn{2}{c}{Peak position}  &
\colhead{Flux density$^a$} &  \colhead{Angular size} \\
\cline{2-3}
\colhead{}                  & \colhead{$\alpha$(2000)}      &
\colhead{$\delta$(2000)}    & \colhead{(mJy)}  &
\colhead{(\arcsec)} \\
}
\startdata
18470-0044 A & 18 49 37.78 & -00 41 01.8 &  2.11$\pm$0.16  & $1.2\times0.9$ \\ 
18470-0044 B & 18 49 35.94 & -00 41 05.9 &  2.45$\pm$0.14  & $1.5\times1.3$ \\ 
19217+1651   & 19 23 58.82 &  16 57 41.2 &  50.4$\pm$0.16  & $0.6\times0.3$ \\ 
23151+5912   & 23 17 20.90 &  59 28 47.8 &  1.70$\pm$0.19  &    $\leq1$     \\ 
\enddata
\tablenotetext{a}{Flux densities are corrected for the primary beam
response. The flux density of IRAS 19217+1651 is the sum from components A and B.}
\end{deluxetable}

\begin{deluxetable}{lcccccc}
\tablewidth{0pt}
\tablecolumns{6}
\tablecaption{DERIVED PARAMETERS \label{tbl-derived}}
\tablehead{
\colhead{Source} & \colhead{D} & \colhead{Diameter} &
 \colhead{EM} &  \colhead{n$_e$}  & \colhead{N$_i$} & \colhead{S.T.} \\
 \colhead{} & \colhead{(kpc)} & \colhead{(pc)} &
 \colhead{(pc~cm$^{-6}$)}  & \colhead{(cm$^{-3})$} & \colhead{(s$^{-1}$)} & 
 \colhead{} \\
}
\startdata
18470-0044$-A$ & 8.2  & 0.042 & $1.1\times10^6$ & $4.3\times10^3$ & $2.1\times10^{46}$ & B0.5\\ 
18470-0044$-B$ & 8.2  & 0.057 & $5.3\times10^5$ & $2.5\times10^3$ & $1.5\times10^{46}$ & B0.5\\ 
19217+1651$-A$ & 10.5 & 0.009 & $7.0\times10^8$ & $2.5\times10^5$ & $5.5\times10^{47}$ & O9.5\\ 
19217+1651$-B$ & 10.5 & 0.041 & $4.3\times10^6$ & $8.4\times10^3$ & $6.3\times10^{46}$ & B0\\  
23151+5912     &  5.7 & 0.0006$^a$ & $2.3\times10^9$ & $1.7\times10^6$ & $6.2\times10^{45}$ & B0.5\\
\enddata
\tablenotetext{a}{The size of this source is estimated from the
fit to its spectrum and not from an observed size.}
\end{deluxetable}

\vfill\eject



\begin{figure}
\epsscale{1.0}
\plotone{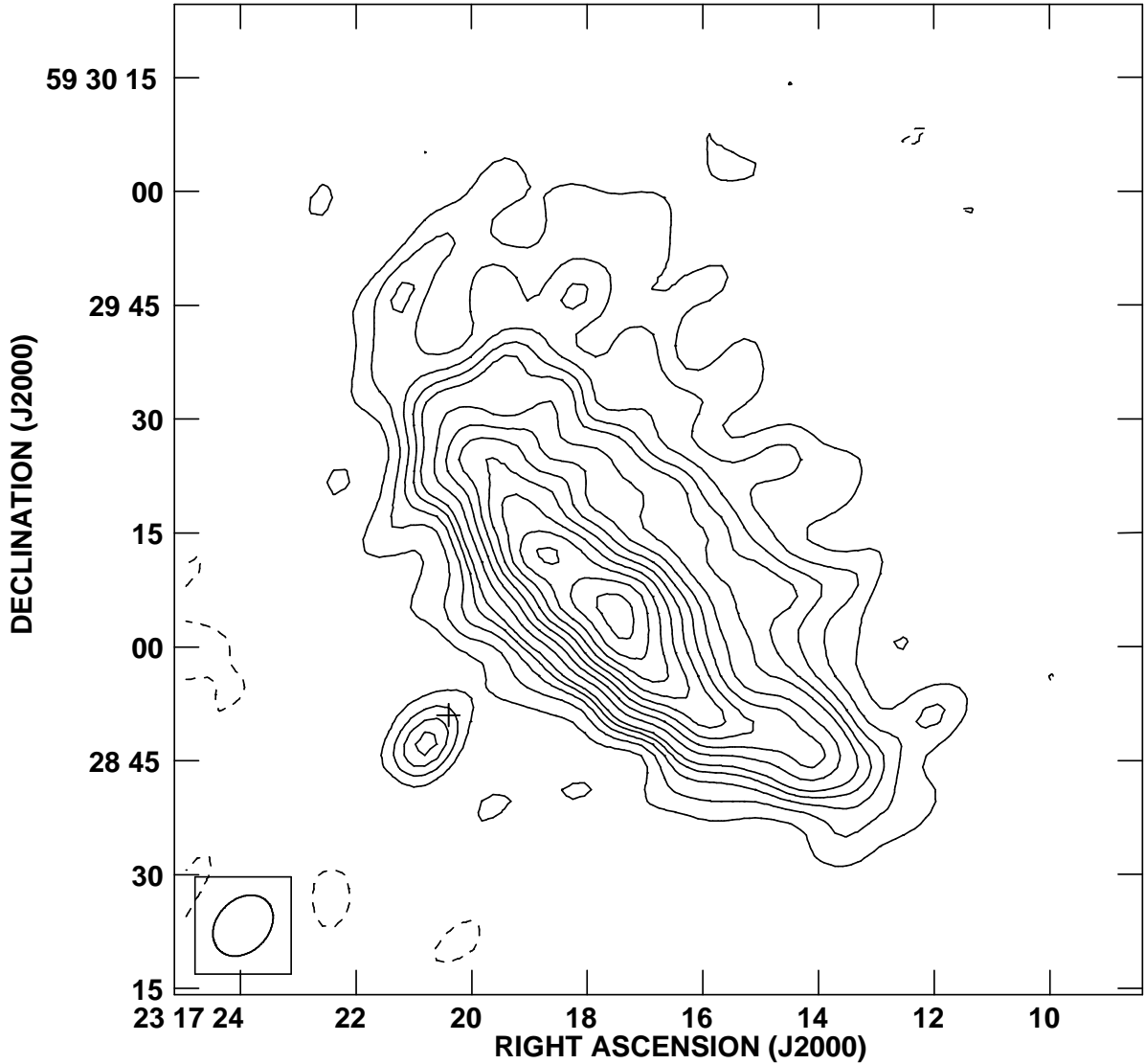}
\vspace{-5.0mm}
\caption
{VLA radio continuum 8.44 GHz image towards IRAS 23151+5912.
The beam is shown in the lower left corner.
Contour levels are -1, 1, 2, 3, 4, 5, 7, 9, 11, 13, 15, 17, and
21 $\times$ 0.060
mJy per beam. Beam is $9.0\times6.8$ \arcsec.
The cross indicates the peak position of the dust core (Williams et al. 2004).
\label{fig-23151xband}}
\end{figure}


\begin{references}

\reference{}


\reference{} Araya, E., Hofner, P., Sewilo, M., Goss, W.M., Linz, H., Kurtz, S., 
 Olmi, L., Churchwell, E.,  Rodr\'{\i}guez, L.F., \& Garay, G. 2007, ApJ,
 submitted

\reference{} Avalos, M., Lizano, S., Rodr\'\i guez, L.F., Franco-Hern\'andez, R.,\&
     Moran, J.M. 2006, ApJ, 641, 406
 
\reference{} Beuther, H., Schilke, P., Menten, K.M., Motte, F., Sridharan, T.K.,
 \& Wyrowski, F. 2002a, ApJ, 566, 945

\reference{} Beuther, H., Walsh, A., Schilke, P., Sridharan, T.K., Menten, K.M.,
  \& Wyrowski, F. 2002b, A\&A 390, 289

\reference{} Bonnell, I.A., Bate, M.R., \& Zinnecker, H. 1998, MNRAS, 298, 93

\reference{} Cesaroni, R., Felli, M., Testi, L., Walmsley, C.M., \& Olmi, L. 1997,
A\&A, 325, 725

\reference{} De Pree, C.G., Rodr\'\i guez, L.F., \& Goss, W.M., 1995, \rmaap, 31, 39

\reference{} Devine, D., Bally, J., Reipurth, B., Shepherd, D., \& Watson, A. 1999,
AJ, 117, 2919

\reference{} Garay, G., \& Lizano, S. 1999, \pasp, 111, 1049

\reference{} Garay, G.,  Brooks, K., Mardones, D., Norris, R.P., \& Burton, M.G.
  2002, \apj, 579, 678

\reference{} Garay, G., Mardones, D., Bronfman, L., Brooks, K.J.  Rodr\'\i guez,
   L.F., G\"usten, R., Nyman, L-{\AA}, Franco-Hern\'andez, R., \& Moran, J.M.  2007, 
   A\&A 463, 217

\reference{} G\'omez, L., Rodr\'\i guez, L.F., Loinard, L., Lizano, S., 
  Poveda, A. \&  Allen, C. 2005, \apj, 635, 1166

\reference{} Gonz\'alez-Avil\'es, M., Lizano, S., \& Raga, A.C. 2005, \apj, 621, 359

\reference{} Hoare, M.G., Kurtz, S.E., Lizano, S., Keto, E. \& Hofner, P. 2007,
  in Protostars and Planets V, ed. B. Reipurth, D. Jewitt, \&  K. Keil, 
  University of Arizona Press, Tucson, 181 

\reference{} Hollenbach, D. J. 1997, in Herbig-Haro Flows and the Birth of Low-Mass 
  Stars, ed. B. Reipurth \& C. Bertout (Dordrecht: Kluwer), 181

\reference{} Keto, E. 2002, \apj, 580, 980

\reference{} Keto, E. 2003, \apj, 599, 1196

\reference{} Kurtz, S.E. 2000, R.M.A.A (Conf. Series), 9, 169

\reference{} Lada, C.J. 1991, in The Physics of Star Formation and Early Stellar
Evolution, ed. C.J. Lada \& N.D. Kylafis, (Dordrecht: Kluwer), 329

\reference{} Mart\'\i , F., Rodr\'\i guez, L.F., \& Reipurth, B. 1993,
ApJ, 416, 208

\reference{} Mezger, P.G., \& Henderson, A.P. 1967, \apj, 147, 471

\reference{} Portegies Zwart, S.F., Makino, J., McMillan, S.L.W., \& Hut, P.
1999, A\&A, 348, 117

\reference{} Rodr\'\i guez, L.F., Garay, G., Curiel, S., Ram\'\i rez, S.,
Torrelles, J.M., G\'omez, Y., \& Vel\'azquez, A. 1994, ApJ, 430, L65

\reference{} Rodr\'\i guez, L.F., Garay, G., Brooks, K.J., \& Mardones, D. 2005,
 ApJ, 626, 953
 
\reference{} Shepherd, D.S., Claussen, M.J., \& Kurtz, S.E. 2001, Science, 292, 1513

\reference{} Shepherd, D.S., Watson, A.M., Sargent, A.I., \& Churchwell, E. 1998,
ApJ, 507, 861

\reference{} Shu, F.H., Adams, F.C., \& Lizano, S. 1987, ARAA, 25, 23

\reference{} Shu, F.H, Najita, J., Galli, D., Ostriker, E., \& Lizano, S. 1993,
in Protostars and Planets III, eds. E.H. Levy \& J.I. Lunine, 3

\reference{} Sridharan, T. K., Beuther, H., Schilke, P., Menten, M. \& Wyrowski, F., 
   2002, ApJ, 566, 931

\reference{} Williams, S.J., Fuller, G.A., \& Sridharan, T.K. 2004, \aap, 417, 1152

\reference{} Xie, T.,  Mundy, L.G., Vogel, S.N., \&  Hofner, P. 1996, 
 ApJ, 473, L131

\reference{} Yorke, H.W. \& Welz, A. 1996, A\&A, 315, 555

\reference{} Zhang , Q., Hunter, T.R., Sridharan, T.K., Ho, P.T.P. 2002, ApJ,
566, 982

\end{references}
\end{document}